Modeling interactions of narrowband large amplitude whistler-mode waves with electrons in the solar wind inside ~.3 AU and at 1 AU using a particle tracing code

Cynthia Cattell and Tien Vo
School of Physics and Astronomy, University of Minnesota, Minneapolis, MN 55455

ABSTRACT: The discovery of large-amplitude narrowband whistler-mode waves at frequencies of tenths of the electron cyclotron frequency in large numbers both inside ~.3 AU and at ~1 AU provides an answer to longstanding questions about scattering and energization of solar wind electrons. The waves can have rapid nonlinear interactions with electrons over a broad energy range. Counter-propagation between electrons and waves is not required for resonance with the obliquely propagating waves in contrast to the case for parallel propagation. Using a full 3d particle tracing code, we have examined interactions of electrons with energies from 0 eV to 2 keV with whistler-mode waves with amplitudes of 20 mV/m and propagation angles from 0 to 180 degrees to the background magnetic field. Interactions with wave packets and single waves are both modeled based on observations at ~.3 AU and 1 AU. The simulations demonstrate the key role played by these waves in rapid scattering and energization of electrons. Results provide evidence for nonlinear effects, indicating that quasi-linear methods are not adequate for modeling the role of whistlers in the evolution of solar wind electrons. Strong scattering and energization for some initial energy and pitch angle ranges occurs for both counter-propagating and obliquely propagating waves. The strong scattering of strahl electrons counteracts the pitch angle narrowing due to conservation of the first adiabatic invariant as electrons propagate away from the sun into regions of smaller magnetic field. Scattering also produces the hotter isotropic halo. The concomitant limiting of the electron heat flux is also relevant in other astrophysical settings.

1. Introduction

Many researchers have studied the evolution of solar wind electron distributions as they propagate away from the Sun. If only adiabatic effects are included, the field-aligned suprathermal strahl electrons would narrow in pitch angle with distance from the Sun, losing perpendicular energy as the magnetic field decreases to conserve the first adiabatic invariant. Instead, satellite observations from ~.2 to >5 AU have shown that the pitch angle width increases with radial distance (Maksimovic et al., 2005; Štverák et al., 2009; Halekas et al., 2020a), and that strahl may be completely scattered by ~5.5 AU (Graham et al., 2017). Because the strahl electrons carry the bulk of the heat flux, many studies are framed as determining the mechanisms that control the heat flux (Gary et al., 1994; Bale et al., 2013; Halekas et al., 2020b). Studies of strahl evolution and heat flux control have assessed the relative roles of Coulomb collisions and wave particle interactions, often concluding that the wave particle interactions are necessary (Phillips and Gosling, 1990; Vocks, 2012; Bale et al., 2013; Boldyrev and Horaites, 2019).

Whistler-mode waves have frequently been invoked as a plausible mechanism to scatter the strahl because interactions with whistler mode waves do not conserve the first adiabatic invariant since the wave frequency and electron gyrofrequency are comparable (Schulz and Lanzerotti, 1974). Until recently most theoretical studies focused on waves propagating parallel to the solar wind magnetic field. For these waves, the resonance condition, $\omega - \vec{k} \cdot \vec{v_e} = n\Omega_e$, can only be satisfied if the whistler-mode waves propagate sunward, opposite to the bulk of the electrons (Vocks et al., 2005; Gary and Saito, 2007). If waves propagate anti-sunward, they can interact with only the small portion of the electrons that travel sunward.

Studies utilizing STEREO waveform capture data revealed the existence of large amplitude narrowband waves (NBWM) at frequencies of ~0.2 $f_{ce}$ (electron cyclotron frequency) that propagate at large angles to the magnetic field (~60–65 degrees) (Breneman et al., 2010; Cattell et al., 2020). The waves have electric field amplitudes ranging from 10s to up to >100 mV/m: ~1 to 3 orders of magnitude larger than previously observed in the solar wind, with parallel components as much as 30% of the perpendicular component. For oblique waves, the resonance condition can be met for electrons and waves propagating in the same direction; thus the observations of large amplitude oblique waves opened up a more significant role for whistler-mode waves. Similar NBWM have been observed inside 0.3 AU by Parker Solar Probe (PSP) (Agapitov et al., 2020;

Cattell et al., 2021a) with more variable wave angles, including some wave packets that propagated sunward. Strahl electrons are strongly scattered by these waves over a broad energy range from ~100 ev to ~1 keV (Cattell et al., 2021b). Two recent particle-in-cell (PIC) simulations found that oblique whistlers were excited by heat flux and strongly scattered electrons, one in the context of solar flares (Roberg-Clark et al., 2019) and one in the solar wind near .3 AU (Micera et al., 2020).

In this report, we focus on the interaction of solar wind electrons with whistler-mode waves using a 3d particle tracing code with initial electron distributions and whistler properties based on those observed both at ~1 AU and inside ~.3 AU enabling us to compare scattering in these two different regions. In section 2, we show an example of a coronal mass ejection with the waves of interest, and briefly describe the particle tracing code. Section 3 shows the results obtained for interaction of electrons with energies from 0 eV to 2 keV, for both wave packets and a single wave. Section 4 discusses comparisons to previous studies and the implications of the results.

1. Whistler-mode waves and particle tracing study

An example of the association of the large-amplitude oblique whistler-mode waves with solar wind structures at 1 AU is shown in Figure 1, which plots a coronal mass ejection on March 24-25, 2017, identified by Jian et al.(2006) in data from STEREO-A (Luhmann et al., 2008; Bougeret et al., 2008). The leading shock can be seen in the jump in the magnetic field in panel a and density in panel b. Vertical blue lines are less coherent whistler-mode waves (see Cattell et al., 2020) and vertical gold lines are narrowband whistler mode waves (examples in the bottom panels). The four example 2.1s waveform capture electric fields, with one component perpendicular to the magnetic field in red, and the parallel component in blue, show that the waves contain significant parallel electric fields, and that the waves occur in packets with durations of a few tenths of a second. Cattell et al. (2020) show a stream interaction region with similar close-packing of whistlers, and statistics including correlations to electron properties.

Inside ~.3 AU, PSP data show that the narrowband whistlers are also often seen for many hours, usually occur in regions of smaller variable background magnetic fields, and sometimes in association with magnetic field switchbacks (Agapitov et al., 2020; Cattell et al., 2021a, 2021b). Propagation angles range from near parallel, sometimes sunward propagating, to highly oblique. Average dB/B are ~.05 with values up to >.12.

Utilizing a full 3-d particle tracing code, with background magnetic fields, densities, and whistler-mode parameters based on the STEREO and PSP observations, we have examined the response of core, halo and strahl electrons to the waves. Electrons with initial energies from 0 eV to 2 keV, pitch angles from 0° to 180°, and gyrophases of 0° to 360° are traced. For the .3 AU packet case, electrons up to 4 keV were traced to allow a direct comparison to the PIC simulations of Micera et al. (2020). Weighting of results is performed using electron parameters based on Wilson III et al. (2019) for 1 AU, and on Halekas et al. (2020a) for inside .3 AU.

The simulations utilize a 3d relativistic test particle code based on that of Roth et al. (1999) and Cattell et al. (2008), modified for solar wind magnetic field and density conditions. Results of an early version of the code (Breneman et al.; 2010) showed prompt electron scattering, as much as 30 to 40 degrees in <0.1s, with energy changes also occurring. The current program is adapted for vectorized calculations of a distribution of test particles interacting with an input wave. The whistler waves are modeled in two different ways: (1) a single wave with a fixed frequency (0.15 $f_{ce}$) and wave angle; and (2) a wave packet formed from 11 waves of varying frequencies. Wavelengths are determined from the cold plasma dispersion relation for a uniform plasma. In the simulation, we assume a uniform background field $\vec{B} = B_0 \hat{z}$ and follow the motion of the electrons under the combined influence of the background magnetic field and the whistler-mode waves. The use of the uniform background magnetic field is justified because the interactions are rapid. In addition, the electrons making the largest excursions along the magnetic field travel distances over which the curvature of the field due to the Parker spiral is negligible.

In the energy range of interest, some particles cross multiple resonances in a short period of time (see Figure 2), as indicated by the resonance mismatch $\nu = \frac{\gamma}{\omega_{ce}}(\omega - k_\parallel v_\parallel)$ (Roth et al., 1999). Due to the short-duration and frequent resonant interactions, the electrons often go through stiff regions in the differential equations describing their motion, requiring more computational resources to solve numerically. Thus, instead of using the standard 4th order Runge-Kutta integrator, we employ the Boris algorithm described in Ripperda et. al. (2018). Solutions calculated using the Boris method are known to be uniformly bounded in energy error (Qin et. al., 2013), and thus able to calculate particle dynamics with great accuracy over a long simulation period. Additionally, we perform a variational calculation of the Lyapunov exponents at each time iteration to ensure the local phase space volume around the particles' trajectory is conserved to ensure the Boris algorithm's efficiency. Our code and the Hamiltonian analysis of the interactions are described in more detail in Vo et al. (2021).

We will show simulation results for two sets of parameters, one for 1 AU ($B_0$=10 nT and n=5 cm$^{-3}$ based on Cattell et al., 2020) and one for ~.2–.3 AU ($B_0$=50 nT and n=300 cm$^{-3}$ based on Cattell et al., 2021a). For the single wave cases, $\frac{f}{f_{ce}}$=0.15, the electric field amplitude is 21 mV/m, and we use three wave angles with respect to the background magnetic field, 5°, 65° and 175°. For the wave packets, $\frac{f}{f_{ce}}$=0.15 is the center frequency and the wave angles are 0°, 65° and 180°.

Examples of the interactions are shown in Figure 2, which plots the time series for three different electrons interacting with a single wave propagating at 65° using the 1 AU parameters. From top to bottom the initial electron properties are: (a) kinetic energy of 100 eV and pitch angle of 0°; (b) kinetic energy of 1000 eV and pitch angle of 0°; and (c) kinetic energy of 1000 eV and pitch angle of 180°. For each case, the panels plot (1) the resonance mismatch, $v$, (2) the relativistic kinetic energy W (in eV), and (3) the pitch angle. Time is normalized to wave periods. For clarity, only the first ~10 wave periods are shown although simulations were run to ~60 wave periods. In the 100 eV, initial pitch angle of 0° case(panels a), when the electron crosses the resonance line ($v=0$), the pitch angle increases to >100°; after subsequent interactions, the pitch angle averages ~100°, and the energy is decreased to an average of ~40 eV. In the 1000 eV, initial pitch angle of 0° case (panels b), the electron crosses the resonance line ($v=0$) multiple times with anti-correlated jumps in energy (~300–350 eV) and pitch angle (~80° to >100°), settling at ~700 eV and ~100°. Trapping of the electrons is indicated by the regions where $v$ remains close to zero. The 1000 eV electron with initial pitch angle of 180° (panels c) crosses the resonance at t~3.5, with a rapid increase in energy to ~1400 eV. After many other resonant interactions, the average energy is ~1300 eV and the average pitch angle is ~50°. The resonance mismatch reaches higher values as the electron initial energy increases, consistent with the operation of higher order resonances. In all three cases electrons are strongly scattered at the resonance crossing, and the change in pitch angle is anticorrelated with the change in pitch angle as expected from theoretical models of scattering (Brice, 1964; Kennel and Petschek,1966; and Albert, 2009). A detailed understanding of the interactions requires examination of resonance broadening and overlap for the specific wave parameters (Karimabadi et al., 1990, 1992), which is addressed in Vo et al. (2021).

The results for the full distributions for 1 AU single wave cases after ~60 wave periods are presented in Figure 3a. All four panels show, from left to right, the core, halo, strahl and total distribution, with the same color bars. The top panel shows the initial distribution, the second panel shows the results for a wave angle of 5° (strahl velocity and wave phase velocity in the same direction), the third panel the results for a wave angle of 175° (strahl velocity and wave phase velocity in opposite directions), and the fourth panel shows the results for a wave angle of 65°, all at the simulation end time of ~60 wave periods. The white 'x's' indicate the locations where the n=1,0 and -1 resonances cross the $v_\perp$=0 axis (see Roberg-Clark et al., 2019), and the black circles plot constant energy curves in the wave frame. Note that the electrons plotted in the core, halo and strahl panels indicate the electrons that were initially in that category. As expected, the strahl electrons are significantly more scattered by the 175° wave than by the 5° wave, and the total distribution contains significantly fewer anti-sunward moving electrons. In the case of the 65° wave, the scattering of the strahl is more symmetric. There is reduced flux in the parallel (anti-sunward) direction, and broad flux peaks at pitch angles around 30°. In addition, the core is more strongly heated in the interaction with oblique waves. In comparing the scattering observed for the 5° and 175° waves to that seen for the 65° wave, it is important to note that the wave amplitude of 20 mV/m, based on observations of oblique waves, was used for all wave angles. Observations at 1 AU, however, have found only small amplitude (<1 mV/m) waves at parallel angles (Lacombe et al. 2014; Stansby et al., 2016; Tong et al., 2019). The final distributions shown for the 5° and 175° waves, therefore, greatly overestimate the potential impact of parallel propagating waves at ~1 AU.

The comparable results for single waves inside .3 AU are shown in Figure 3b (same format and same color bar as Fig. 3a). Because the halo is only a very small component of the electron population, only core and strahl are included in the initial distributions (based on Halekas et al., 2020a). Another key difference is that large amplitude parallel propagating waves and sunward propagating waves are observed inside .3 AU; therefore, the results for 5° and 175° waves represent possible scattering processes. Some features are seen at both radial distances, including the decrease in fluxes along the magnetic field for all wave angles and the clear energy dependent cutoff in the scattering beyond 90° pitch angles in the 175° wave angle case. For the .3 AU parameters, the core electrons are heated, especially for the 65° wave, and population centroid moves to the location of where the n=0 (Landau resonance) intersects the $v_\perp$=0 axis. For the 65° wave, there is strong scattering beyond 90° pitch angle along the constant energy circle in the wave frame at ~2 keV. The energy dependence of the scattering is clearly seen in the .3 AU figures; similar features were seen in the PSP observations (Cattell et al., 2021b).

The role of whistler waves is more accurately modeled using a group of wave packets, as can be seen in the examples in Figure 1. Figure 4 (in the same format as Figure 3) plots the results of the interactions with wave packets centered at 0° and 65° for 1 AU, and 0°, 65° and 180° for .3 AU. Comparing the final distributions for the single wave cases to the packet cases indicates that the interaction with a single wave results in more energization and/or scattering; however, some similar features can be seen. For the near parallel cases, both have peaks in the flux at an angle off parallel; for the packet, this is seen in both the negative and positive direction. The parallel packet case has constant energy (in the wave frame) enhancements at high energies for Vz>0 (clearly visible in the strahl panel), which extend to lower energies in the packet case as would be expected due to the fact that there are multiple resonant energies in the packet. For the oblique wave cases, both show minima in the magnetic field-aligned and anti-field aligned directions, most clearly for the single wave case. Both the 0° and 65° packets have a region of enhanced flux at a constant parallel velocity (most obvious in the higher density core). In addition, the 65° packet and single wave cases have very prominent 'horn'-like elements at the higher energies; the 0° waves show these 'horns,' most clearly in the 1 AU case. These structures are comparable to the multiple 'horns' described by Roberg-Clark et al. (2019) and attributed to higher order resonances. Similar features were seen at early times in the simulations of Micera et al. (2020) when oblique whistlers were growing. The 180° packet case was only run for .3 AU because no sunward propagating whistlers have been found at 1 AU. The 180° packet shows energy dependent features similar to the 175° single wave, but electrons are not scattered as significantly. The most significant scattering of the strahl and heating of the core is observed in interactions with the oblique waves. The evolution of these and other features can be clearly seen in the movies in the Supplementary Material, which plot the distributions versus time.

To more clearly illuminate the role of trapping in different resonances, as well as scattering and energization for different initial pitch angles and energies, we utilized multiple time series videos examining specific initial conditions, such as ones looking at restricted energy ranges and pitch angles or initial propagation directions. These visualizations reveal the very complex interactions of electrons with large amplitude waves. Trapping and de-trapping of electrons in different resonances occurs. The amount of scattering and energization is very dependent on the initial electron energy and pitch angles, and the wave angle and whether it is a single wave or packets.

As discussed in the introduction, simple arguments based on the resonance conditions imply that electrons propagating in the same direction as a parallel propagating wave will not interact strongly with the wave. The simulations show, however, that by ~6 wave periods some of the highest energy (~1300 eV) electrons near 90° pitch angle are scattered so that their parallel velocity component is anti-sunward, and these electrons interact strongly with the wave. The interaction with the wave packets occurs more quickly for the electrons that are initially moving anti-parallel to the wave; the highest energy electrons already scattered to all pitch angles by 3 wave periods, whereas, in the parallel case, this occurs by ~ 5 wave periods. For the 65° wave packet, the interaction with the electrons traveling in either direction occurs within a time comparable to that for the antiparallel electrons in the parallel propagating wave packet case (~3 wave periods).

As expected for the 0° packet, the electrons with initial pitch angles near 180° are rapidly scattered. Some electrons are trapped in the n=-1 resonance: for initial energies of ~300 eV, the trapping is at pitch angles of ~50° with energization up to ~700 eV; and, for initial energies ~1000 eV, the trapping is at ~60°–70° with energization to ~1500 eV. At the highest energies, the electrons are experience alternate trapping and de-trapping. In contrast, the 0° electrons lose energy and are only weakly scattered by the end of the simulation. Some electrons initially at ~90° and ~1000 eV are trapped in n=0 and the n=-1 resonances. During the interaction with the 65° packets, electrons with all initial conditions are very rapidly scattered, with the response of the 180° particles slightly preceding those at 0° at all energies. There is evidence for trapping of some electrons in all three resonances N=-1,0, +1). The scattering results in a loss of particles with pitch angles along the magnetic field.

4. Discussion and Conclusions

Results of a fully 3d relativistic particle tracing code with wave properties based on the narrowband whistler mode waves observed by at ~1 AU by STEREO (Cattell et al., 2020) and inside ~.3 AU by PSP (Cattell et al., 2021a), and initial electron distributions based on Wind data at 1 AU (Wilson III et al., 2019) and the PSP data inside .3 AU (Halekas et al., 2020a) show that solar wind electrons from core through strahl (energies from a few eV to a few keV) can be strongly scattered and/or energized. The whistlers scatter the strahl to produce the isotropic halo, as required to explain the observed strahl width and the changes in the relative densities of the strahl and halo with distance from the Sun, and the limitation of the electron heat flux. The conclusions are significant not only for the solar wind, but potentially also for other high beta astrophysical settings including the interstellar medium and inter-cluster medium.

Studies of electron acceleration both in the solar wind (Saito and Gary, 2007a,b) and in the Earth's radiation belts (Tao et al., 2013) have demonstrated that the wave packet structure dramatically affects the acceleration and scattering of electrons. Comparison of our results for the single wave to those for wave packets indicate that, although some features are weakened in the packet cases, there are new scattering features observed only with packets. This is due to the existence of multiple resonances and resonant overlaps associated with the different frequencies in the packets.

The simulations using parameters appropriate for inside ~.3 AU show that both the highly oblique and the sunward propagating parallel whistler-mode waves rapidly scatter the strahl to produce the hotter more isotropic halo, and to reduce the heat flux carried by the strahl electrons. Oblique waves more significantly scatter the strahl, consistent with the study of Halekas et al. (2020b) which showed that the electron heat flux was constrained by the oblique heat flux fan instability. Features seen in the simulations, including peaks at angles to the magnetic field in some energies, scattering past 90°, energy dependent scattering and evidence for higher order resonances, are consistent with the direct observations of electron scattering and energization by narrowband whistler mode waves using the PSP wave and electron observations (Cattell et al., 2021b). The core heating seen in our simulations is also seen when the large amplitude narrowband whistlers are observed (Cattell et al., 2021a), but not in studies of core electron temperature that are not constrained to times with waves. In addition, results are consistent with observed changes in strahl and halo made by both PSP and Helios (Maksimovic et al., 2005; Halekas et al., 2020a, b).

We can compare our results with two PIC simulations that examined the interaction of whistlers with electrons. Roberg-Clark et al. (2019) modeled this process in the context of solar flares, with an energetic outflowing electron kappa distribution and a cold return current population. Large amplitude oblique whistlers, propagating first with and then against the heat flux, and electron acoustic waves were excited. The whistler power peaked at highly oblique angles with amplitudes comparable to those observed in the PSP and STEREO data. The electrons were rapidly scattered with multiple energy dependent 'horn'-like features similar to those we observed in the particle tracing for the oblique wave case at .3 AU. The simulations of Micera et al.(2020) were initialized with core and strahl distributions based on the same PSP measurements (Halekas et al., 2020a) that we utilized. They found that initially highly oblique whistler waves were excited, and scattered the strahl to produce a single 'horn'-like feature. At later times, parallel whistlers were excited, producing additional scattering that further isotropized the electron distributions. Both PIC studies discuss the time evolution of the role different resonances and the importance of nonlinear interactions with the large amplitude waves.

The changes seen in the electron distributions for the 1 AU simulations are most directly applicable to understanding the observed radial evolution around 1 AU. Štverák et al. (2008) showed that, in the slow solar wind (defined as <500 km/s), the high energy tails of both the halo and strahl become more pronounced with radial distance, with a large increase around 1 AU for halo. The NBWM highly oblique whistlers observed by STEREO at 1 AU occur primarily in association with the slow wind as defined by Štverák et al. (Cattell et al., 2021a); thus, the fact that the simulations show the development of high energy tails is consistent with these observations. As discussed in the introduction, a clear strahl is often absent outside 1 AU. This is also consistent with our simulation results that show that scattering by oblique waves with properties based on those observed at 1 AU results in an almost isotropic distribution.

In summary, the results of our particle tracing simulations with parameters based on observations of electrons and waves inside .3 AU and at 1 AU provide strong evidence for the central role of oblique whistler-mode waves in the evolution of solar wind electrons. The whistler scattering of the strahl electrons produces the halo and limits the electron heat flux. The nonlinear interactions, including trapping, clearly indicate that a quasi-linear approach cannot accurately model the scattering process. Our conclusions are also applicable to other high beta astrophysical plasmas.

Figures

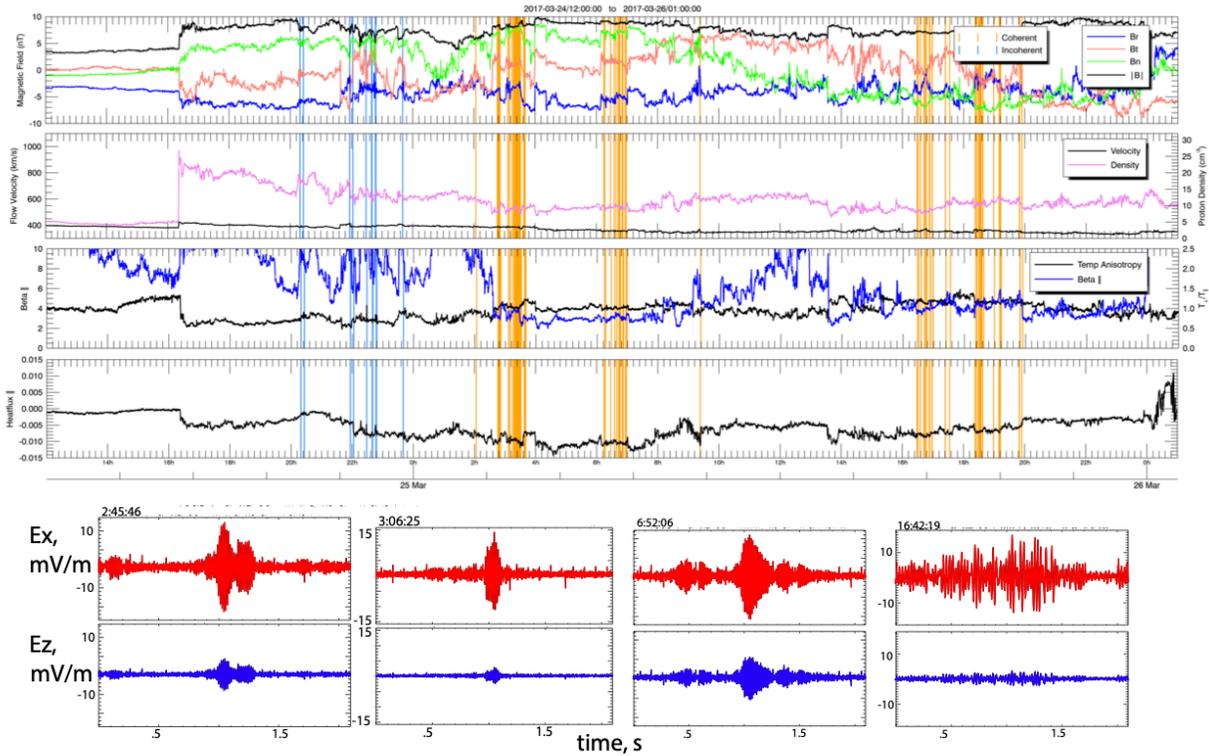

Figure 1. Example of a CME with whistler-mode waves. (a) Magnetic field in RTN coordinates and magnitude; (b) solar wind velocity (black) and density (pink); (c) electron temperature anisotropy (black) and beta parallel (blue); (d) parallel electron heat flux. Vertical blue lines are less coherent whistler-mode waves and vertical gold lines are narrowband whistler mode waves (examples bottom panels). Bottom: Example 2.1s waveform capture electric field, one perpendicular to the magnetic field component in red, parallel component in blue.

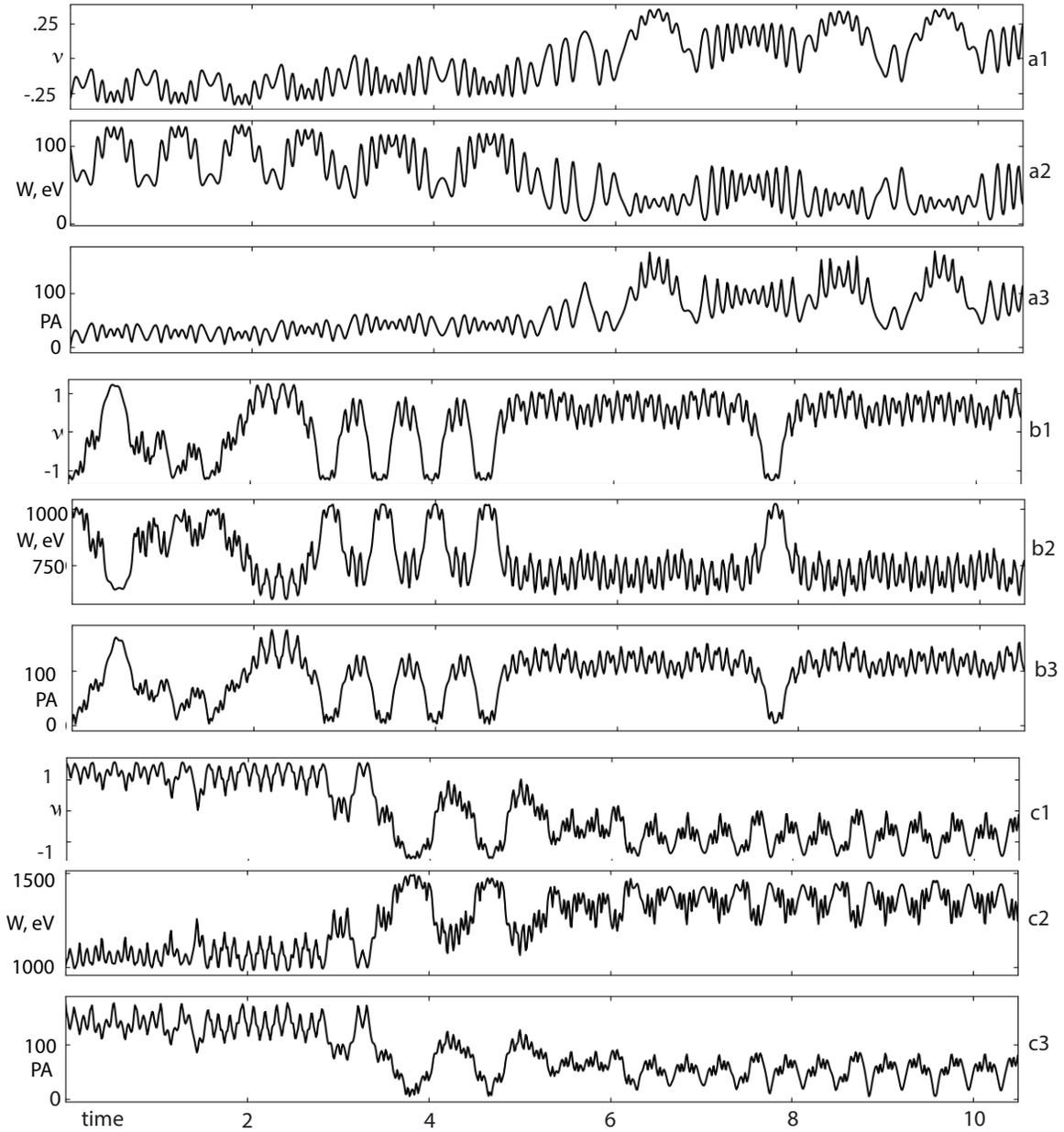

Figure 2. Time series of for the interaction of an electron with a single wave propagating at 65° for three different electron initial conditions: (a) initial kinetic energy of 100 eV and initial pitch angle of 0°; (b) Initial energy of 1000 eV and pitch angle 0°; and (c) Initial energy of 1000 eV and pitch angle 180°. Time is normalized to wave periods. For each set of initial conditions the panels are from top to bottom: (1) the resonance mismatch; (2) the relativistic kinetic energy W; and (3) the pitch angle.

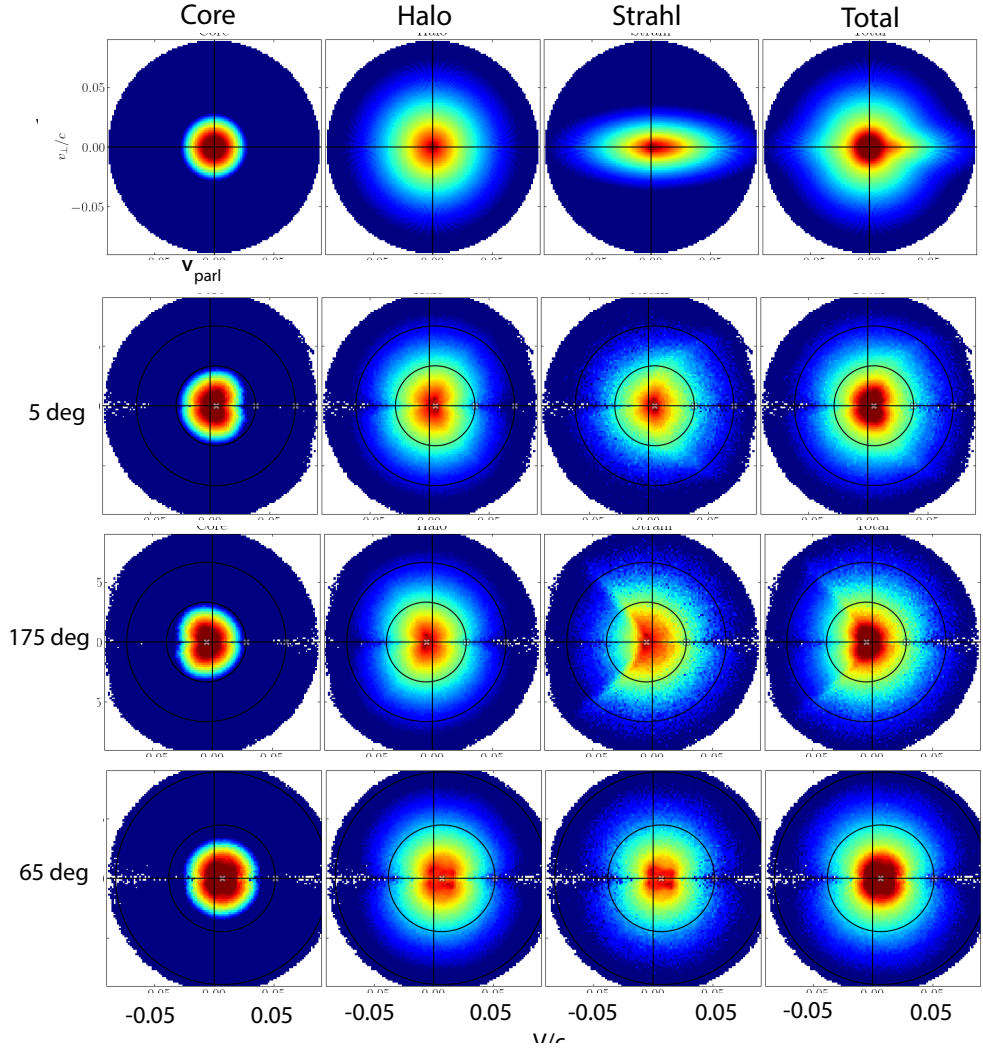

Figure 3. (a) Interaction of solar wind electron distributions at 1 AU with single whistler wave. Top panels show the initial distributions. The next panels show the distributions after interacting with an almost parallel (in the direction of strahl), anti-parallel (opposed to the strahl), and very oblique wave for 60 wave periods. The white x's indicate the location where the n=-1,0, and 1 resonances intersect $v_\perp = 0$. The black circle plot constant energy in the wave frame.

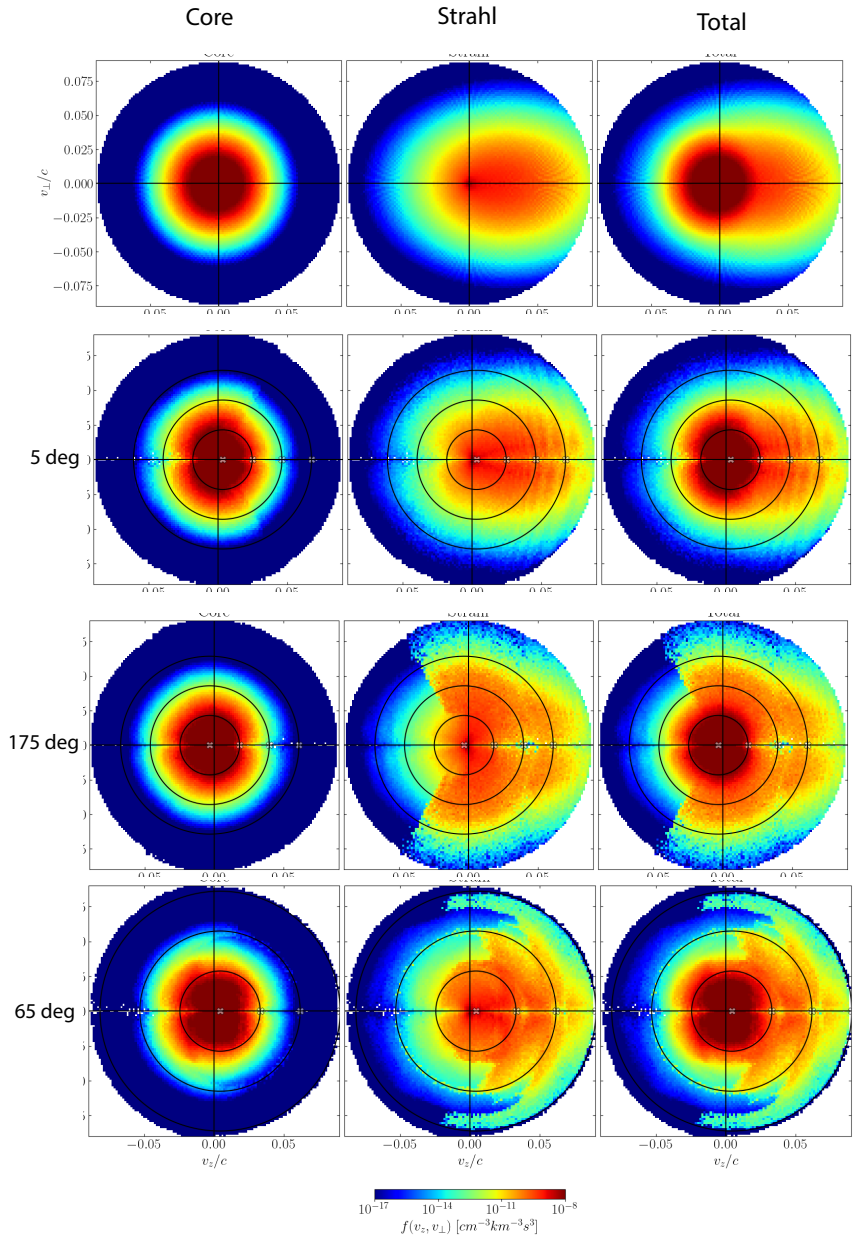

Figure 3. (b) Interaction of solar wind electron distributions at 0.3 AU with single whistler wave. The top panel shows the initial distributions for core, strahl and total distributions at t=0. The next panels show the distributions after interacting with an almost parallel (in the direction of strahl), anti-parallel (opposed to the strahl), and very oblique wave for 60 wave periods. The white x's indicate the location where the n=-1,0, and 1 resonances intersect $v_\perp = 0$. The black circles plot constant energy in the wave frame.

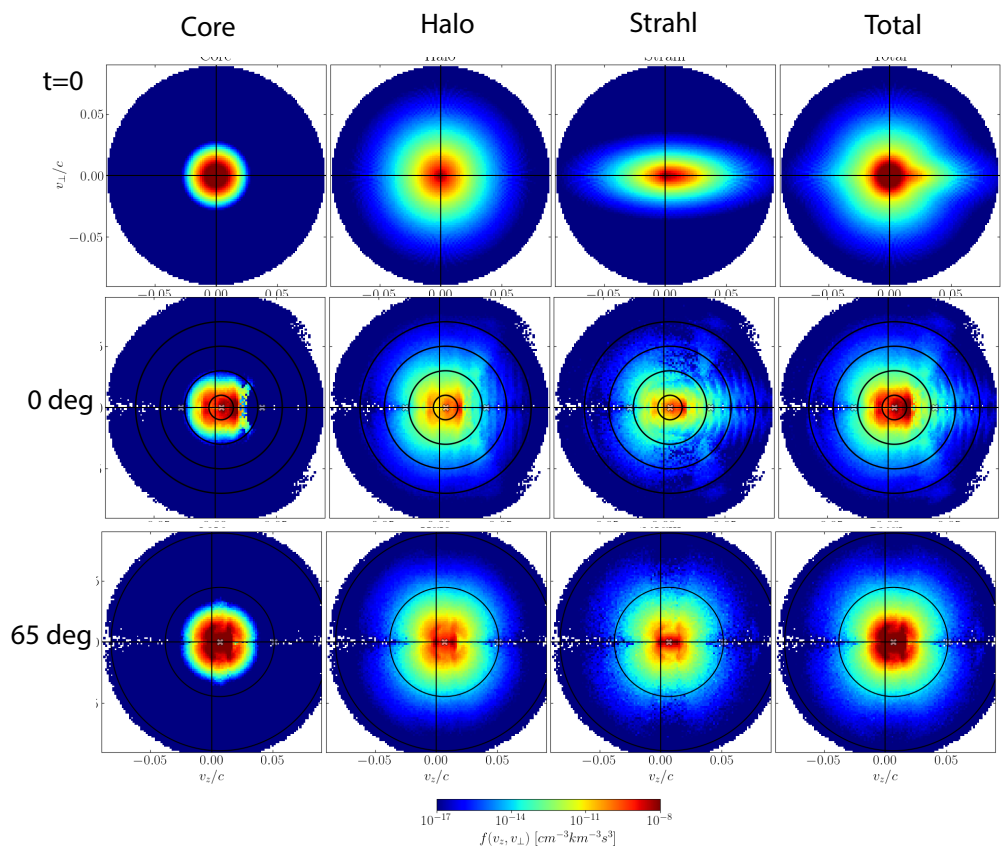

Figure 4a. Interaction of electrons with wave packets (same format as Figure 3)a.

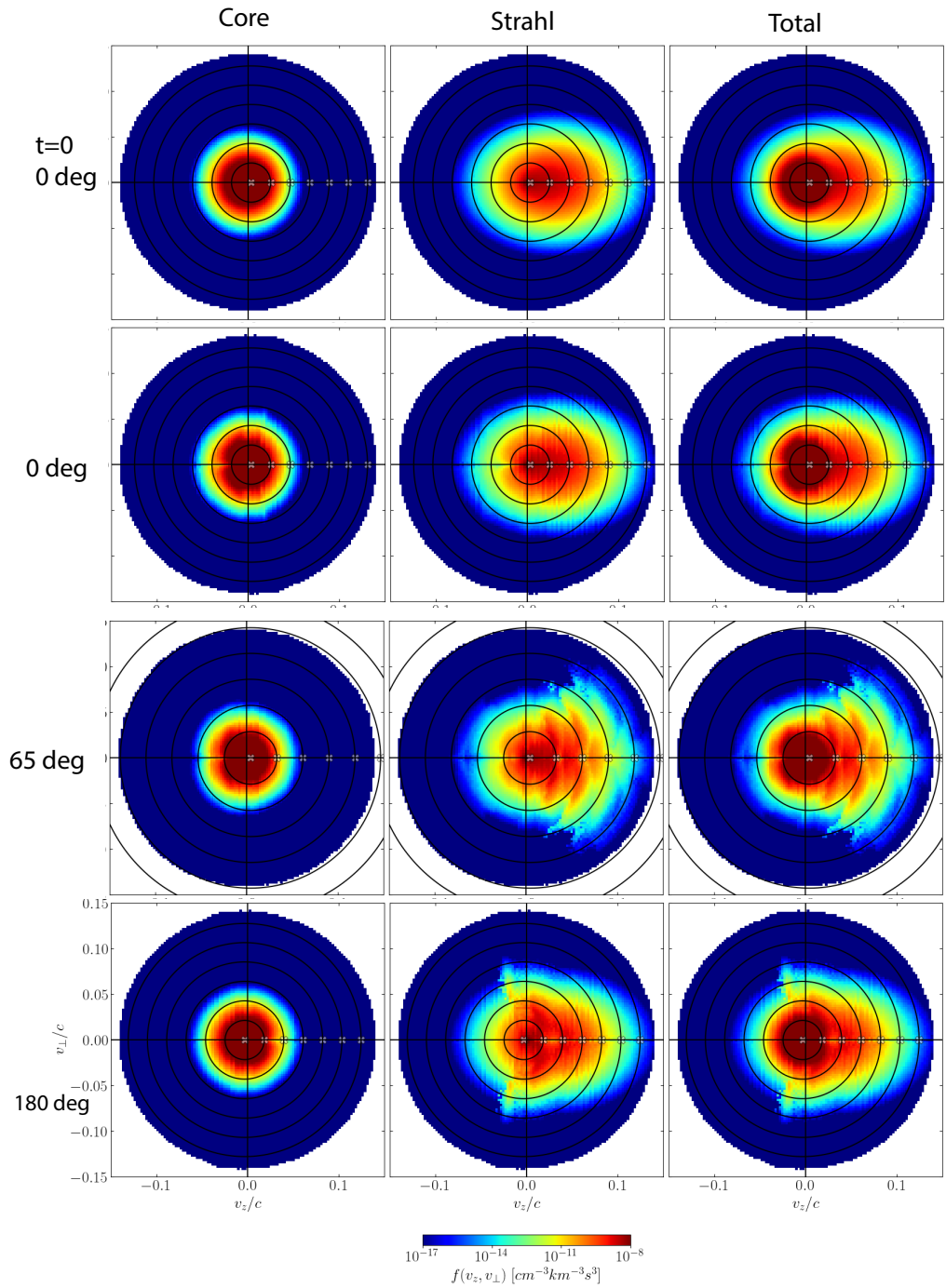

Figure 4b. Interaction of wave packets with electrons at ~.3 AU. Final time is 45 wave periods. Same format at Fig. 3b

Acknowledgements: We thank M. K. Hudson, R. Lysak, T. Jones and V. Roytershteyn for helpful discussions and comments. The authors acknowledge the Minnesota Supercomputing Institute (MSI) at the University of Minnesota for providing resources that contributed to the research results reported within this paper. URL: http://www.msi.umn.edu. This work was supported by NASA grants NNX16AF80G, 80NSSC19K305, and NNX14AK73G.